\documentclass[twocolumn,showpacs,amsmath,amssymb,prl]{revtex4-1}

\usepackage{graphicx} 
\usepackage{bm} 
\usepackage{epsfig}
\usepackage{color}

\begin{document}

\title{Geometrical guidance and trapping transition of human sperm cells\\}

\author{A. Guidobaldi$^1$, Y. Jeyaram$^2$, I. Berdakin$^3$, V. V. Moshchalkov$^2$, C. A. Condat$^3$, V. I. Marconi$^3$, L. Giojalas$^1$, A. V. Silhanek$^4$}

\affiliation{$^1$IIByT, CONICET, UNC, C\'ordoba, Argentina.\\
$^2$Institute for Nanoscale Physics and Chemistry, KU Leuven, B--3001 Leuven, Belgium.\\
$^3$Facultad de Matem\'atica, Astronom\'ia  y F\'isica, Universidad Nacional de C\'ordoba and IFEG-CONICET, X5000HUA C\'ordoba, Argentina.\\
$^4$D\'epartement de Physique, Universit\'e de Li\`ege, B-4000 Sart Tilman, Belgium.\\}

\begin{abstract}

The guidance of human sperm cells under confinement in quasi 2D microchambers  is investigated using a purely physical method to control their distribution. Transport property measurements and simulations are performed with dilute sperm populations, for which effects of geometrical guidance and concentration  are studied in detail.  In particular, a trapping transition at convex angular wall features is identified and analyzed. We also show that highly efficient microratchets  can be fabricated by using curved asymmetric obstacles to take advantage of the spermatozoa specific swimming strategy.
\end{abstract}


\maketitle

Understanding sperm dynamics under confining microgeometries  is a general problem and a major challenge both from the  basic biophysics and the complex fluids points of view. It is also crucial for microfluidics and biomedical control applications. Our knowledge of the swimming cell motilities in  unbounded media cannot be directly extrapolated to their behavior in complex environments such as those found in the oviduct or in the lab-on-a-chip microfluidic devices used to control and analyze small samples or for in-vitro reproduction procedures. In these cases, the characteristic length scales are of the same order as the cell size, i.e. a few micrometers. Under these circumstances, confined self-propelled microorganisms undergo substantial changes in their locomotion habits, adapting their dynamics to intricate porous media or to solid surfaces vicinity~\citep{Mino2011}, reducing their speed close to boundaries~\citep{Frymier1995} or adjusting their morphology and motility in very narrow channels~\citep{Mannik2009}.

\begin{figure}[ht]
\begin{center}
\includegraphics[angle=0, width=7.5cm]{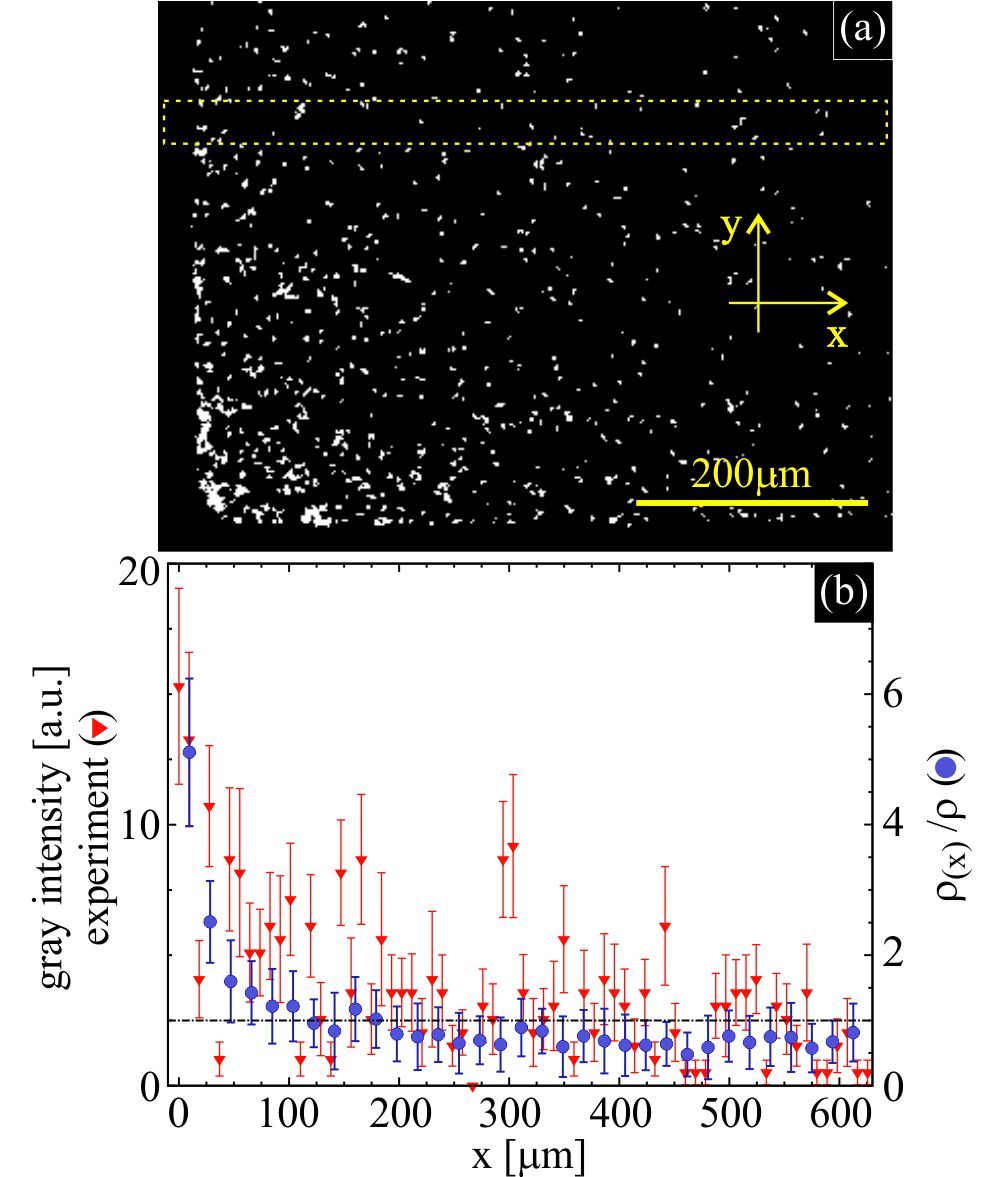}
\caption{Color online. (a) Microscopy image of a two dimensional sperm distribution close to the chamber corner. Depth $L_z=20\mu m$. The walls of the container become apparent by the accumulation of sperm cells. (b) Quantification of the sperm distribution proportional to the gray intensity. Experimental data is indicated with triangles and numerical results with circles. The dashed  line corresponds to the average concentration.} \label{fig:wall}
\end{center}
\end{figure}

It has been shown that  microswimmers with very different propulsion systems are similarly attracted to the walls and to swim parallel to the surface ~\cite{Rothschild1963,Winet1984,Frymier1995,Cosson2003,Woolley2003,Berke2008,Garcia2008,Li2009,Drescher2011}.  It is believed that this attractive force has hydrodynamic origin although other possible mechanisms have been proposed ~\citep{Smith2011, Gaffney2011, Elgeti2011}. Several models have been introduced to describe the swimming along surfaces (see Ref.~\citep{Dunstan2012} and references therein). Interestingly, the direct observation of the cell-wall attraction (see Fig.~\ref{fig:wall}) have led to the design of ratchet devices that guide and sort self-propelled cells using asymmetric obstacles~\citep{Galajda2007, Kantsler2013}. In particular, different microfluidic devices have been created to either increase sperm cell quality or enhance their concentration~\citep{Chen2011,Tasoglu2013,Matsuura2013}. The creation of inhomogeneous distributions of swimmer populations via asymmetric obstacles has been shown to be particularly efficient for run-and-tumble bacteria~\citep{Galajda2007, Hulme2008, Angelani2009, Kim2010}. Alternative ways of achieving nonuniform distributions have also been obtained combining symmetric funnels and flux~\citep{ Altshuler2013}. Nowadays, numerous theoretical treatments are available  to account for the effects of asymmetric obstacles on active particles distributions~\citep{Wan2008, Tailleur2009, Berdakin2013a, Berdakin2013b}. Tumbles, rotational diffusion and  collisions  are efficient mechanisms for separating the cells from the surface, thus permitting bacteria to be reinserted into the bulk of the confining microchamber. Although larger and generally faster than bacteria, a similar rectification effect should, in principle, be observed for {\it human sperm cells} and we could take advantage of it in applications. A distinct property, however, is their physiology and specific  swimming strategy. Since spermatozoa do not tumble,  their detachment from the wall, in a very diluted sample, will presumably proceed from other effects, as the observed ciliary beating of bull spermatozoa near corners ~\cite{Kantsler2013} or their rotational diffusion~\cite{Li2009}.

In this Letter, we show experimentally and numerically the importance of the particular behavior of  human sperm cells  near walls and corners.  This is key  to properly design  microfluidic chambers able to achieve efficient cell guidance and controlled inhomogeneous cell population distributions. We show that the asymmetric designs  used to generate inhomogeneous bacterial distributions need to be revisited when dealing with sperm cells. Indeed, V-shaped asymmetric pillar arrays introduced to rectify bacterial displacement~\citep{Galajda2007}  lead to sperm trapping near the obstacle apices. A high concentration of sperm cells is then obtained at every angular wall feature at the expense of emptying the space away from the confining walls. This undesired effect renders the chambers with  V-shaped obstacles not optimal for controlling and directing sperm cells continuously and uniformly, without cell trapping, a typical requirement in biomedical applications. To avoid this trapping mechanism while still keeping the guiding capability of the designs, we introduce rounded and asymmetric $\cup$-shaped funnel arrays and rounded box corners. We demonstrate that, by building a row of $\cup$-shaped funnels facing a row of $\cap$-shaped funnels, it is possible to obtain a high concentration of {\it uniformly distributed} sperm cells in the region confined between the rows. We have achieved the most efficient geometry on the basis of simulations, which allow us to further investigate the dynamics of confined sperm using the motility parameters directly extracted from our experimental sample tracks. This work is then the result of a strong positive feedback among soft-lithography microfabrication, biological measurements and  phenomenological modeling without adjustable parameters.

\begin{figure}[ht]
\begin{center}
\includegraphics[angle=0, width=7.5cm]{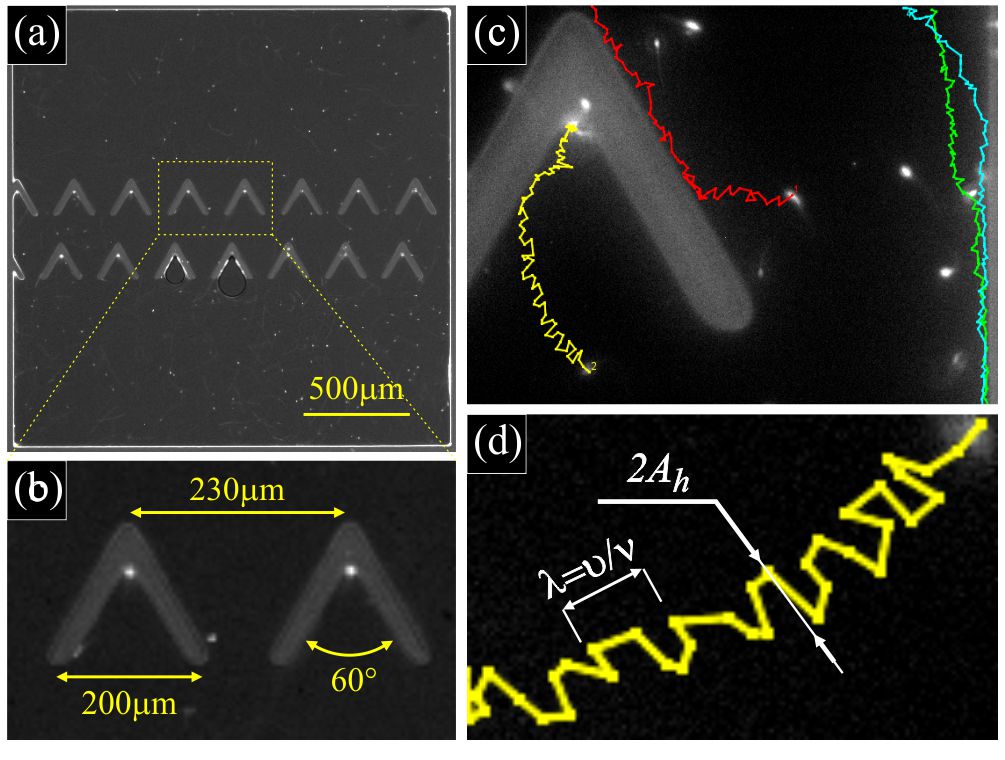}
\caption{Color online. Human sperm trapping near the apices of a V-shaped ratchet. (a) Chamber with a double row of V-shaped obstacles similar to those used for bacterial motion rectification, clearly showing accumulation at all apices.   (b) Geometrical parameters and  detail of the sperm accumulation (brigther dots) at the V-apices. (c) High magnification details of the sperm cells behavior: cell trapping at a convex angle and swimming along walls. Some traked cell trajectories are shown with colored lines. See movie in Ref.~\citep{movie}. (d) Period and amplitude of the cell head oscillation.} \label{fig:v}
\end{center}
\end{figure}


{\em Fabrication of hydrostatic microfluidic chambers.} The quasi 2D nanoliter chambers were prepared in SU-8 or EPOCLAD negative optical resists deposited on glass substrates.  The materials were chosen to ensure (a) optical transparency allowing transmission microscopy imaging, (b) high aspect ratio structures (height/width $>$ 20), (c) biocompatibility, and (d) watertightness. Both epoxies show no particular differences concerning biocompatibility within the time frame of the experiments, although SU-8 provided higher resolution for nanofabrication. We have observed no major differences between both epoxies concerning the physical response of the sperm cells, thus evidencing the robustness of the main results of this work.

{\em Sperm preparation and manipulation.} Spermatozoa were separated from the seminal plasma by a discontinuous Percoll gradient \citep{Aitken1988}. The highly motile sperm population was adjusted to 10$^7$ cells/ml with HAM-F10 medium containing 25 mM Hepes and L-glutamine, supplemented with 1\% human albumin. Then, the sperm were loaded with 1 $\mu$M of Fluo4-AM for at least 15 min and then kept in an incubator at $37^\circ$C with 5\% CO2 in air until their use. The chambers were loaded with 0.5 $\mu$l of sperm suspension over the well and covered by sliding a handmade 3$\times$3 mm$^2$ coverslip to avoid air bubble formation. To achieve a seal, excess liquid was absorbed with paper. The edges were covered with mineral oil to prevent air entry. The sperm movement was recorded by fluorescent videomicroscopy using a digital camera connected to and inverted  microscope (Nikon, USA). Recordings of tracks were performed at 25 Hz with the BR Nis Elements software and the image analysis was made with ImageJ free software.

{\em Model.}  The dynamics of $N$ self-propelled swimmers confined to a micro-patterned $L_xL_y$ two-dimensional box is simulated by taking into account the observed head oscillations (see Fig.~\ref{fig:v}(d)). Each swimmer, under a flagellar self-propelling force and viscous friction, and free of inertia,   is represented by a soft disk of radius $R_h$. Its swimming velocity is given by:
  \begin{equation}\label{eq:eq_dinam}
    \begin{split}
      &\dot{x}_i(t) = v_i cos(\varphi_i) + A_h \omega \cos(\omega t) \sin(\varphi_i)\\
      &\dot{y}_i(t) = v_i sin(\varphi_i) - A_h \omega \cos(\omega t) \cos(\varphi_i),
    \end{split}
  \end{equation}

with $v_i$ being the speed of swimmer $i$, which is taken from a Gaussian distribution to represent the non-uniform cell population, $\varphi_i$  the angle between its instantaneous direction of motion and the $x$-axis, $A_h$ the head oscillation amplitude, $\nu$ the oscillation frequency and $\omega=2\pi\nu$. Since we take the constant parameters, $A_h$, $\omega$, and $v$, as the averages of the measured values of the experimentally observed head tracks, {\em we have no fitting parameters}.  The average direction of motion changes due to asymmetries of the propulsion system and thermal fluctuations. These changes are quantified by a rotational diffusion coefficient, $D_r$, a parameter obtained from experiments by averaging over the 94 longest tracks, selected from a set of 558 analyzed trajectories. Then $\varphi_i$ changes due to $D_r$ as $\Delta \varphi = \sqrt{2D_r \Delta t} \chi$ with $\chi$ a Gaussian random variable of unit variance and $\Delta t$ the integration time step. In this phenomenological manner it is possible to induce a deviation and a consequent separation of the cells swimming along the walls.

The interaction among swimmers and swimmer-wall are represented as follows: When a swimmer with velocity $v_i$ collides with another  swimmer, a reversion of the velocity components parallel to the line joining the swimmer centers is established. The repulsion between swimmers $i$ and $j$ arises when $r_{ij}$ $=$ $\lvert \vec{r}_i - \vec{r}_j \rvert$ $\leq$ $R_h^i + R_h^j$. Similarly, if the distance between swimmer $i$ and wall $k$, $r_{ik}$ $\leq$ $r_h^i + w/2$, where $w$ is the width of the wall, a velocity $v_i$ normal to the wall moves the swimmer away from it. Immediately after the collision, the swimmer rotates to head parallel (antiparallel) to the wall if $\hat{e}_h \cdot \hat{e}_p^k$ is $>$ ($<$) $0$, where $\hat{e}_p^k$ is a unit vector parallel to the wall. The stochastic equations of motion  were integrated with first-order methods.

We used the following parameters: $R_h$ $=$ $2.5$ $\pm$ $0.1$ $\mu m$, $v = 28$ $\pm$ $4$ $\mu m s^{-1}$, $A_h = 1.5$ $\mu m$, $\nu = 10$ Hz, and $D_r$= $0.01$ rad$^2$/s. All the simulation results were averages over 50-100 initial conditions on the velocity distributions of the population. Cells trajectories, average densities in the lower chamber (easy direction of motion), $\rho_e$, and in the upper chamber (hard direction), $\rho_h$,  and rectification effects were measured, with the rectification defined as $r$=$\rho_e/(\rho_e+\rho_h)$. We also investigated the detailed dependence of the results on the shape and number of the obstacles.

\begin{figure}[ht]
\begin{center}
\includegraphics[angle=0, width=8.5cm]{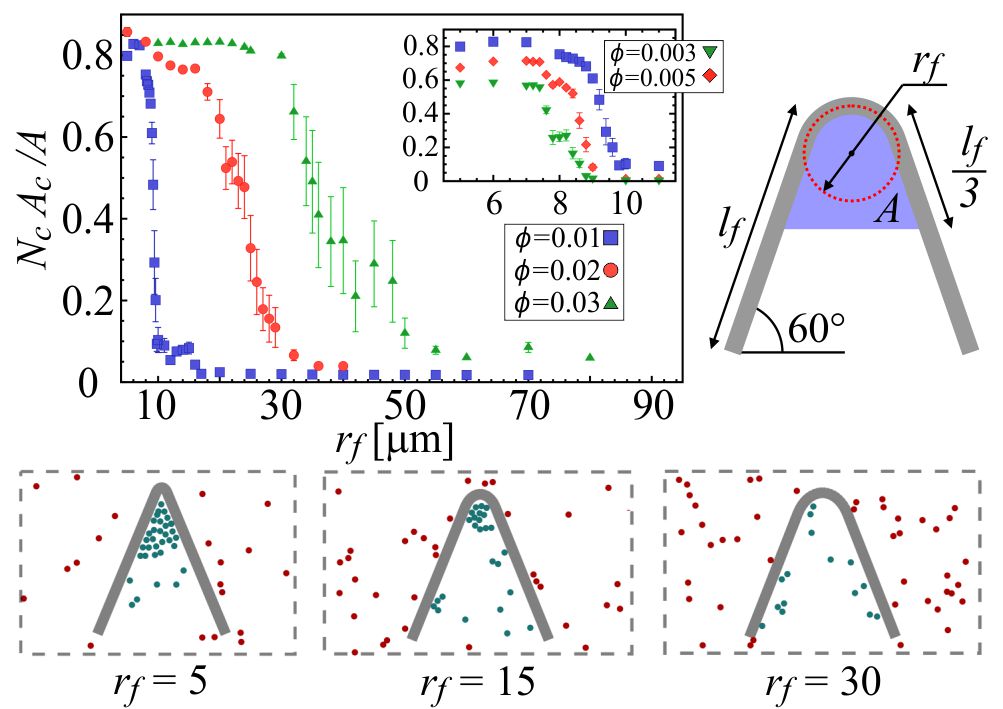}
\caption{Color online. Numerical results of the obstacle curvature effects. Main panel: fraction of the shaded area $A$ in the scheme on the right occupied by sperm cells vs. the radius of curvature, for the indicated sperm density $\phi$. Inset: results at low densities. Lower panels: Snapshots for $\phi = 0.02$ and different  $r_f$  values. $N_c$ is the average number of cells found in $A$. A density of $\phi = 0.01$ corresponds to a cell concentration of $\rho = 6 \times 10^{6}$ cells/ml.} \label{fig:corner}
\end{center}
\end{figure}

{\em  Sperm accumulation near walls.} Spermatozoa were loaded into a well of area 2$\times$2 mm$^2$ and depth 20 $\mu$m, allowing them to swim in an effectively two-dimensional region . In Fig.~\ref{fig:wall}(a) we show a snapshot of one  chamber corner in the {\it xy}-plane. We observe that sperm cells tend to accumulate near the boundaries while maintaining a fairly homogeneous distribution far from them. In order to quantify more precisely the spatial dependence of the cell distribution, we have considered five non-overlapping rectangular boxes such as the one delimited by a dashed line in Fig.~\ref{fig:wall}(a). These boxes, with randomly chosen locations in the region between 250 $\mu$m and 600 $\mu$m above the corner shown in the figure, were divided into vertical bins of 10 pixels width ($\sim$ 18 $\mu$m). We then averaged the intensity over each bin and plotted the result in Fig.~\ref{fig:wall}(b). Clearly, a monotonic increment of the cell density is observed as we approach the wall, reaching a density of about 5 times that in the middle of the box. We remark that the cell accumulation near the walls  was measured in the 2D $xy$-plane and may not necessarily follow the spatial dependence reported using  sperm of other mammals on a drop of fluid compressed between two glass slides~\citep{Li2009}. On the contrary, in our case the wall height, $L_z$ is just $\sim$20 $\mu$m, i.e. about twice the largest dimension of the sperm head. In Fig.~\ref{fig:wall}(b) we also plot the numerical results obtained using experimental values characterizing cell motility (see Fig. ~\ref{fig:v}(d)). Note the good qualitative agreement, given that there are no free parameters. Due to the finite size of the system, the accumulation near the walls generates cell depletion away from them. The cell density decays approximately as the inverse square root of the distance to the wall.

{\em V-shaped obstacles.} The observation that walls attract sperm cells so that they swim close to them encouraged us to design cell sorters following the layouts already tested for bacteria ~\citep{Galajda2007}. One of these designs consisting of a chamber divided by two rows of asymmetric V-shaped obstacles is shown in Fig.~\ref{fig:v}(a). When this chamber is homogeneously inoculated with sperm cells, the cells are guided by the walls, and the asymmetric shape of the funnels leads to a higher population density in the lower chamber. Unfortunately, this design leads to cell directioning but not as efficient as expected due to unwanted cell concentration near the convex angles of the V-shaped obstacles, Fig.~\ref{fig:v}(a-c).
 We used simulations to investigate this phenomenon, keeping the apex angle fixed while changing the rounding (radius of curvature) of the V tip in order to look for a trapping-detrapping transition. The accumulation of  sperm  cells observed near the V-shaped corners and all other corners in the chamber is a direct consequence of their specific swimming strategy, a persistent progressive forward movement
(see the movie in the Supplementary Material \citep{movie}). Our  simulations were performed under periodic boundary conditions for a single V-shaped obstacle as illustrated by the scheme in Fig.~\ref{fig:corner} using various values for the tip curvature. The average sperm surface density $\phi$ is calculated as $\phi = NA_c/(L_xL_y)$, where $A_c$ is the individual cell area and  $L_xL_y$ the total simulated area $L_x=L_y=1000$. The main panel of Fig.~\ref{fig:corner} shows the average occupied area in $A$ as a function of the radius of curvature $r_f$, for three values of $\phi$. Note the sudden detrapping transition obtained at $r_f \sim$ 8 $\mu$m for low densities (inset). That it occurs at 8 $\mu$ m is not surprising if we realize that an individual cell needs a space of the order of $2R_h + 2A_h \sim  8 \mu$m to maneuver out of the corner. Figure ~\ref{fig:corner} exhibits other two properties of the transition. Firstly, it moves to higher values of $r_f$ as we increase the density: since the cells are on the average pushing towards the vertex, increasing their number in the apex region leads to an increase in the jamming probability, hindering detachment. Secondly, it is sharp: If we increment $r_f$ for a given value of $\phi$, there is a $r_f$ value for which one of the cells has room enough to detach from the jam. Detachment of the first cell leaves more room for the rest, facilitating further corner evacuation.

\begin{figure}[ht]
\begin{center}
\includegraphics[angle=0, width=8.5cm]{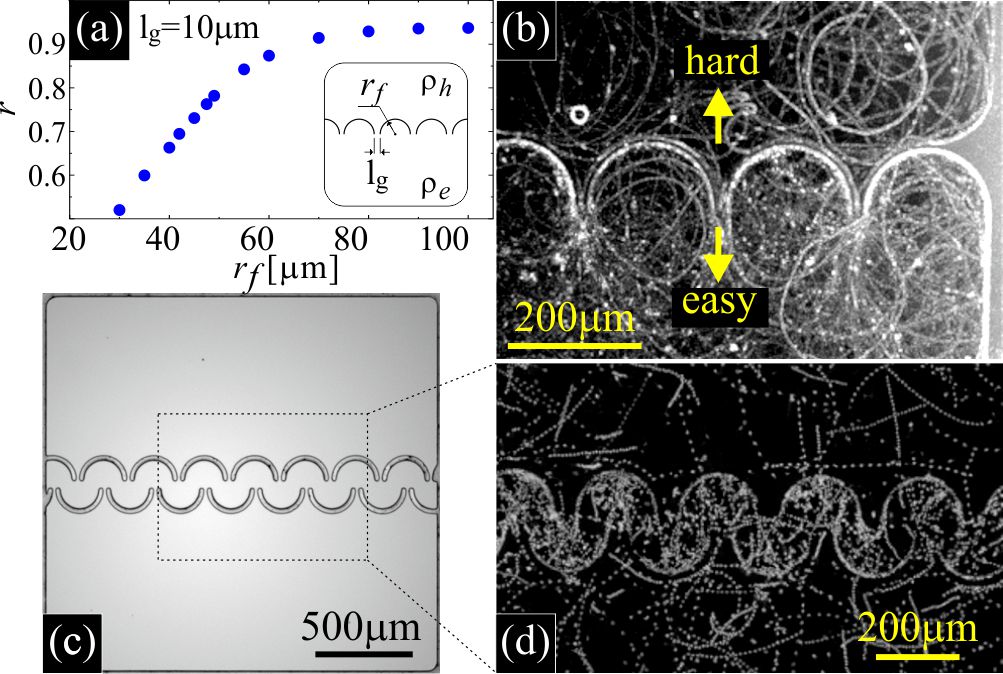}
\caption{(a) Simulated sperm rectification in an 8-gap chamber as a function of the radius of curvature. The rectification is close to 90\% for curvature radii longer than $80\mu m$. Inset: sketch of the simulated ratchet system with rounded obstacles. (b) A single row of $\cap$-shaped funnels guides sperm cells in the easy direction of motion, {\it i.e.} towards the lower half of the chamber. (c) A combination of $\cup$-shaped and $\cap$-shaped funnels was fabricated. (d) The micro-architecture in (c) satisfies our initial goal: a  limited region of highly concentrated sperm cells without corner accumulation.  Numerical and experimental density,  $\rho = 10^{7}$ cells/ml. Microscopy images in (b) and (d) represent time projections evidencing the sperm trajectories.}
 \label{fig:concen}
\end{center}
\end{figure}

\begin{figure}[ht]
\begin{center}
\includegraphics[angle=0, width=7.5cm]{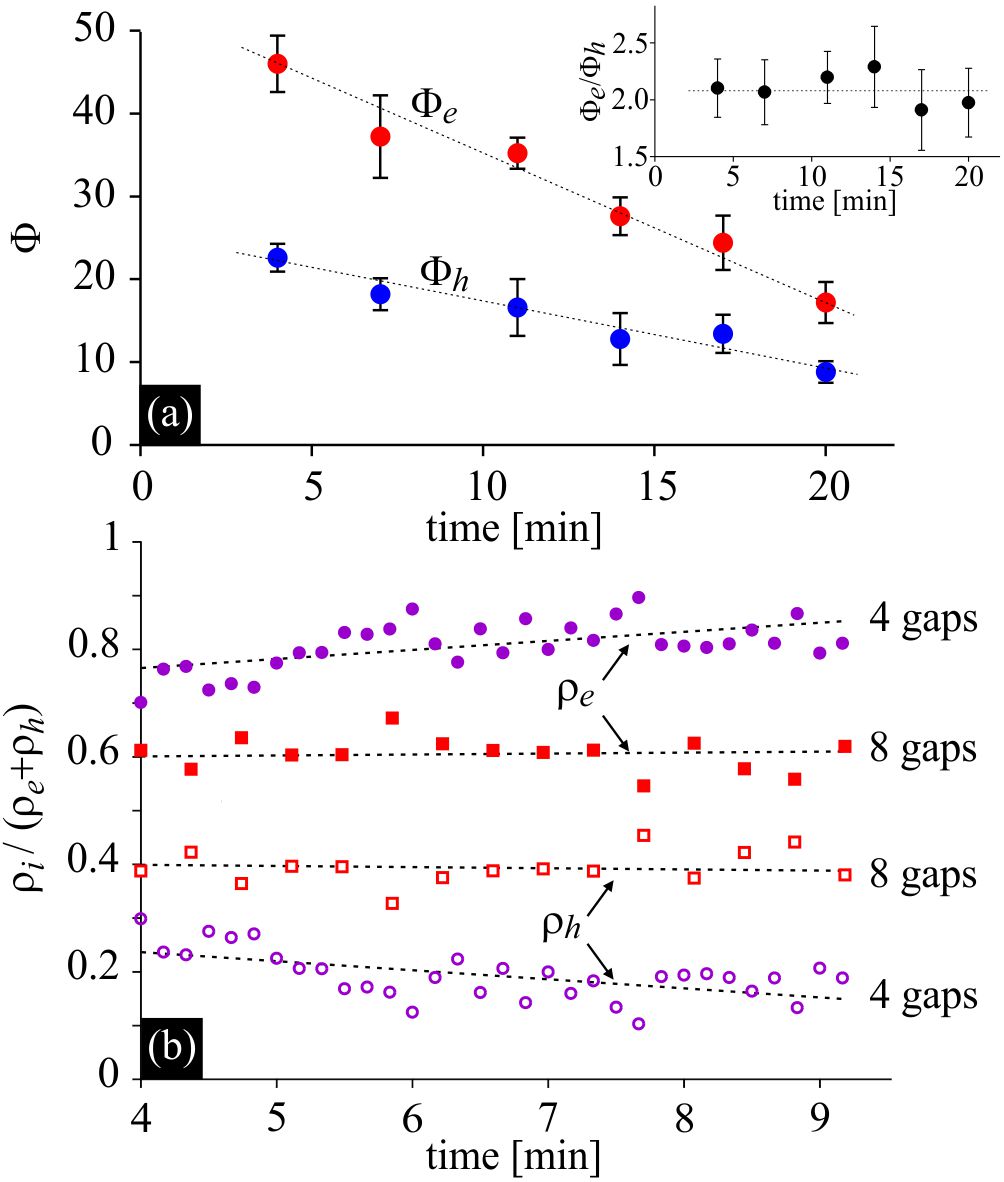}
\caption{Quantifying human sperm rectification. (a) Sperm cell fluxes in the easy ($\Phi_{e}$) and hard ($\Phi_{h}$) ratchet directions {\it vs.} time in the experiment shown in Fig.\ref{fig:concen}(b). $\Phi$ is calculated measuring the number $N_c$ of crossing cells during two minutes. Flux decay is due to cell death (see text). Inset: relative sperm fluxes. (b) Steady state relative concentrations on both sides of the obstacle wall for geometries with 4 and 8 gaps (circles and squares, respectively). The y-axis label $\rho_i$ must be read as $\rho_{e}$ or $\rho_{h}$, as corresponds to the curves labeled as $\rho_i$ inside the panel. Time is measured starting 4 min. after the chamber is closed. Early times are omitted because of experimental constraints.}

\label{fig:flux}
\end{center}
\end{figure}

{\em $\cup$-shaped obstacles.} It is clear from both experiment and simulation that rounding off the tips of the V-shaped obstacles will permit us to reorient the cells towards the bulk of the reservoirs in order to maintain a more homogeneous distribution. We thus designed a chamber where a single row of V-shaped obstacles was replaced by a line of $\cup$-shaped pillars. To optimize the design, we first simulated the cell dynamics in a chamber containing a row of round-shaped funnels (see inset of Fig.~\ref{fig:concen}(a)) with 8 gaps of size $l_g$ = 10 $\mu$m and $w$=25 $\mu$m for several  curvature radius $r_f$, adapting $L_x$ and leaving $L_y$ fixed at 1885 $\mu$m. Figure \ref{fig:concen}(a) exhibits the numerical results, with the rectification $r$ being the normalized steady state cell number in the lower (easy) chamber. The rectification increases in a roughly linear fashion until it saturates at $r_f \sim$ 80 $\mu$m.  The constant cell leakage through the gaps prevents $100\%$ rectification efficiency and instead a  $90\%$ rectification efficiency is observed. Following this template, we fabricated microchambers with a row of round-shaped funnels, as shown in Fig.~\ref{fig:concen}(b), with either  4 or 8 gaps, $r_f \approx$ 100 $\mu$m and $l_g \approx$ 15 $\mu$m. In panel (b) we track individual cell trajectories by projecting the maximum intensity of 100 frames (4 sec.). From this image we see that sperm cells often follow circular paths as pointed out in previous reports on quasi 2D confined systems ~\citep{Cosson2003, Woolley2003, Garcia2008}.

The collected trajectories evidence the existence of an enhanced cell density in the lower chamber, a consequence of the broken funnel symmetry. Furthermore, by using a row of $\cup$-shaped funnels facing a row of $\cap$-shaped funnels as shown in Fig.~\ref{fig:concen}(d), we achieve high sperm concentrations in the space between the rows without corner trapping. These high cell concentrations could be useful for creating a sperm jet or harvesting sperm cells from a reservoir into a narrow channel. The details of the chambers fabrication, such as sharp edges and the sameness in the obstacles, can be appreciated in Fig.~\ref{fig:concen}(c).

We quantified the rectification efficiency by following two different experimental approaches. First we counted the number of cells crossing each gap in Fig.~\ref{fig:concen}(b) in the easy (e) and hard (h) directions within a timeframe of two minutes. After averaging over 5 contiguous gaps, the resulting fluxes, $\Phi_{e}$ and $\Phi_{h}$, are shown in Fig.~\ref{fig:flux}(a) as functions of time. We observe that $\Phi_{e}$ systematically exceeds $\Phi_{h}$, thus showing that the asymmetric structure operates as a  geometrical  ratchet or rectifier, suggesting a purely physical method for controlling sperm cell dynamics and distribution. The linear decay of $\Phi_{e}$ and $\Phi_{h}$ may reflect the death rate of cells in a somewhat hostile 2D confining habitat after 15 min. Irrespective of cell death, we show in the inset of Fig.~\ref{fig:flux}(a) that the ratio of the swimmer fluxes remains practically constant, being twice as likely that a particle moves in the easy direction as in the hard direction. The second approach consists of directly measuring the density of sperm cells in a well-delimited region covering the same area on each side of the chamber. The result of this study is summarized in Fig.~\ref{fig:flux}(b). Interestingly, while a moderate cell separation is observed for an 8-gap chamber, the sorting efficiency is substantially increased if a narrower chamber with only 4 gaps is used, in agreement with the trend of numerical simulations. We believe this to be due to the larger perimeter-to-area ratio of the smaller chamber, which favors easier wall capture.

In summary, we performed original microratchet experiments and numerical studies on human sperm cells. Our simple and minimal phenomenological  model captures the essential observed sperm behavior and gives  excellent predictions without adjustable parameters. We started by verifying that human sperm cells are attracted to boundaries, obtaining the concentration variation near a wall with a very good agreement between experiment and simulation. Surprisingly we found for first time  that sperm cells get stuck near the apices of angular wall features, thus rendering V-shaped structures less suitable for uniform sperm cell directioning and separation.  We then studied the trapping-detrapping transition that occurs when we change the curvature radius of an angular feature, finding that this transition is remarkably sharp and clear. Using our  predicted geometries,  which, incidentally, were in general agreement with very  recent theoretical works ~\citep{Kaiser2012, Deseigne2012},  rows of $\cup$-shaped obstacles were designed and a high sperm cell concentration was achieved, demonstrating that the interaction of the cells with the rounded walls leads to the rectification and control of the sperm movement without trapping the cells. Our non-invasive procedure is an alternative efficient method to achieve sperm sorting ~\citep{Cho2003, Seo2007} and directioning in channels~\citep{Denissenko2012} . We have also shown that a characterization of the specific strategy for swimming along walls of each microswimmer species and of its behavior near corners and along edges, together with a knowledge of the motility parameters under 2D confinement will be crucial for designing and modeling efficient biomedical devices.

{\em Acknowledgments.} We acknowledge financial support from CONICET, MINCyT, and SeCyT-UNC (Argentina), and  the FNRS-CONICET bilateral project (V4/325 C). This work was also partially supported by the FNRS, the Methusalem Funding, and the FWO-Vlaanderen (Belgium).

\end{document}